\documentclass[review]{elsarticle}

\usepackage{bm}
\usepackage{makecell}
\usepackage{multirow}
\usepackage{lineno}
\usepackage{colortbl}
\usepackage{algorithmicx}
\usepackage{algpseudocode}
\usepackage{algorithm}
\usepackage{url}
\usepackage{rotating}
\usepackage{pdflscape}
\usepackage{amsmath}
\usepackage{epstopdf}
\usepackage{arydshln}
\usepackage{amssymb}

\usepackage{graphicx}
\usepackage{overpic}
\usepackage{bm}
\usepackage{multirow}
\usepackage{lineno}
\usepackage{colortbl}
\usepackage{algorithmicx}
\usepackage{algpseudocode}
\usepackage{algorithm}
\usepackage{url}
\usepackage{rotating}
\usepackage{pdflscape}
\usepackage{tikz}

\usepackage{longtable}
\usepackage{arydshln}
\usepackage{caption}
\usepackage{graphicx}
\usepackage[colorinlistoftodos]{todonotes}
\usepackage{algorithm}
\usepackage{algpseudocode}
\usepackage{tikz}

\usetikzlibrary{arrows,positioning}
\usetikzlibrary{plotmarks}
\usetikzlibrary{calc}
\usetikzlibrary{arrows,positioning}
\usetikzlibrary{plotmarks}
\usetikzlibrary{shapes.geometric}

\newcommand{\Ktrans}{K^{trans}}
\newcommand{\ve}{v_e}

\newcommand{\Precision}{\eta_{\rm prec}}
\newcommand{\Recall}{\eta_{\rm tp}}
\newcommand{\Specificity}{\eta_{tn}}

\begin{document}

\begin{frontmatter}

\title{Fully-automated deep learning-powered system for DCE-MRI analysis of brain tumors}

\author{Jakub Nalepa$^{\rm a,b,*}$}
\ead{jakub.nalepa@polsl.pl}
\cortext[mycorrespondingauthor]{Corresponding author}

\author{Pablo Ribalta Lorenzo$^{\rm b}$}
\ead{pribalta@ieee.org}

\author{Michal Marcinkiewicz$^{\rm a}$}
\ead{mmarcinkiewicz@future-processing.com}

\author{Barbara~Bobek-Billewicz$^{\rm c}$}
\ead{bbillewicz@io.gliwice.pl}

\author{Pawel Wawrzyniak$^{\rm c}$}
\ead{pawel.wawrzyniak@io.gliwice.pl}

\author{Maksym Walczak$^{\rm a}$}
\ead{mwalczak@future-processing.com}

\author{Michal Kawulok$^{\rm a,b}$}
\ead{michal.kawulok@polsl.pl}

\author{Wojciech Dudzik$^{\rm a}$}
\ead{wdudzik@future-processing.com}

\author{Grzegorz Mrukwa$^{\rm a}$}
\ead{grukwa@future-processing.com}

\author{Pawel Ulrych$^{\rm c}$}
\ead{pawel.ulrych@io.gliwice.pl}

\author{Michael P. Hayball$^{\rm d}$}
\ead{mike.hayball@fbkmed.co.uk}



\address{$^{\rm a}$Future Processing\\ Bojkowska 37A, 44-100 Gliwice, Poland}

\address{$^{\rm b}$Institute of Informatics, Silesian University of Technology\\ Akademicka 16, 44-100 Gliwice,
Poland, Tel./Fax.: +48 32 237 21 51}

\address{$^{\rm c}$Maria Sklodowska-Curie Memorial Cancer Center and Institute of Oncology\\ Wybrzeze Armii Krajowej 15, 44-102 Gliwice, Poland}

\address{$^{\rm d}$Feedback Medical Ltd.\\Broadway, Bourn, Cambridge CB23 2TA, UK}

\begin{abstract}
Dynamic contrast-enhanced magnetic resonance imaging (DCE-MRI) plays an important role in diagnosis and grading of brain tumor. Although manual DCE biomarker extraction algorithms boost the diagnostic yield of DCE-MRI by providing quantitative information on tumor prognosis and prediction, they are time-consuming and prone to human error. In this paper, we propose a fully-automated, end-to-end system for DCE-MRI analysis of brain tumors. Our deep learning-powered technique does not require any user interaction, it yields reproducible results, and it is rigorously validated against benchmark (BraTS'17 for tumor segmentation, and a test dataset released by the Quantitative Imaging Biomarkers Alliance for the contrast-concentration fitting) and clinical (44 low-grade glioma patients) data. Also, we introduce a cubic model of the vascular input function used for pharmacokinetic modeling which significantly decreases the fitting error when compared with the state of the art, alongside a real-time algorithm for determination of the vascular input region. An extensive experimental study, backed up with statistical tests, showed that our system delivers state-of-the-art results (in terms of segmentation accuracy and contrast-concentration fitting) while requiring less than 3 minutes to process an entire input DCE-MRI study using a single GPU.
\end{abstract}


\end{frontmatter}

\section{Introduction}\label{sec:introduction}

Dynamic contrast-enhanced imaging using magnetic resonance (DCE-MRI) has become a widely utilized clinical tool for non-invasive assessment of the vascular support of various tumors~\cite{RonnebergerFB15}. DCE analysis is performed on a time-series of images acquired following injection of contrast material (tracer) and investigating temporal changes of attenuation in vessels and tissues. Biomarkers extracted from such imaging have been shown to be correlated with physiological and molecular processes which can be observed in tumor angiogenesis (morphologically characterized by an increased number of micro-vessels which are extremely difficult to image directly~\cite{RonnebergerFB15}). Therefore, DCE biomarkers can be used to assess tumor characteristics and stage, and provide an independent indicator of prognosis, enabling risk stratification for patients with cancer~\cite{CUENOD20131187}.

Although DCE biomarkers have been validated against many reference methods and used for assessment of a wide range of tumors, including gliomas, glioblastomas, carcinoids, rectal, renal, and lung tumors, and many others~\cite{CUENOD20131187}, the process of their extraction is fairly time-consuming and prone to user errors as it requires manual segmentation of the MR data. The manual segmentation adversely impacts reproducibility, which is an important issue in clinical applications, particularly for longitudinal studies. In this work, we address these problems and propose a fully-automated approach to assess brain tumor perfusion from DCE-MRI without any user intervention which fits well into clinical practice and can help clinicians decide on an optimal treatment pathway much faster. To the best of our knowledge, such hands-free DCE analysis engines have not been explored in the literature so far.

\begin{figure}[ht!]
	\includegraphics[width=1\columnwidth]{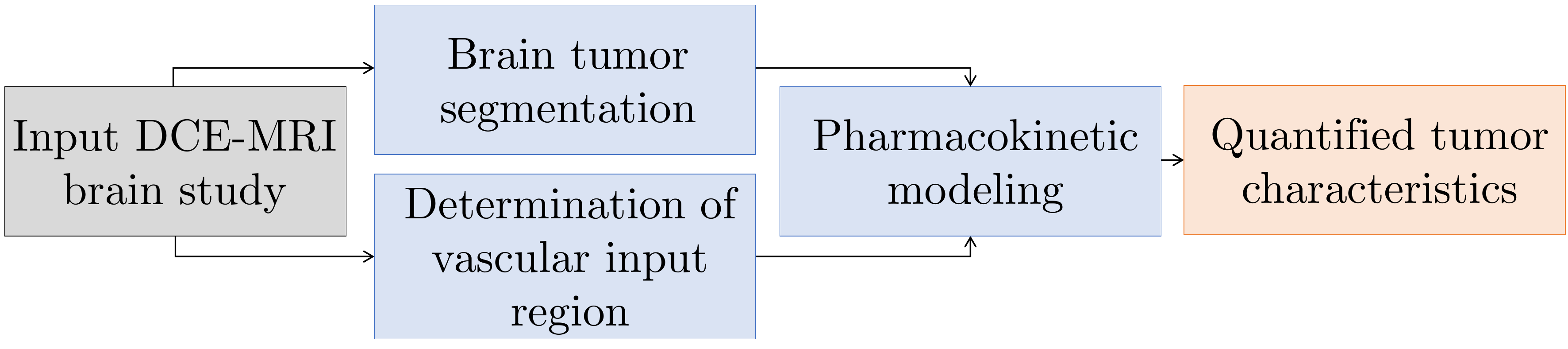}
	\caption[fig:AIFmodels]{Our deep learning-powered approach for assessing brain tumor perfusion does \emph{not} require user intervention---all steps presented in the flowchart are automatic.}
	\label{fig:flowchart}
\end{figure}

The contribution of this work is multi-fold. We propose an extensively validated (Section~\ref{sec:experimental_validation}) system (referred to as ECONIB which unfolds to \textbf{E}nhancing the diagnostic efficiency of dynamic \textbf{C}ontrast-enhanced imaging in personalized \textbf{O}ncology by extracting \textbf{N}ew and \textbf{I}mproved \textbf{B}iomarkers) which requires \emph{no} user intervention for assessing brain tumor perfusion from DCE-MRI data (see its high-level flowchart in Fig.~\ref{fig:flowchart}). Although there were some attempts to automate perfusion analysis in the literature~\cite{doi:10.1038/jcbfm.2010.4}, our system is first-to-date approach for DCE-MRI brain imaging. To make the analysis fully automatic, we introduce:
\begin{itemize}
\item[-] A deep neural network (DNN) for segmenting brain tumors from T2-weighted sequences (Section~\ref{sec:brain_tumor_segmentation}).
\item[-] A real-time image-processing algorithm for determination of the vascular input region from T1 VIBE sequences (Section~\ref{sec:sagittal_sinus_segmentation}). 
\item[-] A new cubic model of the vascular input function used for pharmacokinetic modeling whose aim is to minimize the contrast-concentration fitting error (Section~\ref{sec:pharmacokinetic_modeling}).
\item[-] An end-to-end DCE processing pipeline which can be seamlessly integrated into clinical practice.
\end{itemize}

\section{Related work}\label{sec:related_literature}

The process of extraction of DCE biomarkers from brain MRI requires segmenting brain tumors, determining the vascular input region (most of the state-of-the-art approaches for this task use clustering and intensive 3D processing of brain scans~\cite{Yin2015}---we address their most important shortcoming, being the computational complexity~\cite{PASSAT2007512}, by exploiting very fast image-processing routines coupled with a 3D analysis), and performing the contrast-concentration modeling and analysis. To the best of our knowledge, there are no fully-automated approaches for the DCE analysis proposed in the literature so far, and the existing techniques suffer from lack of reproducibility and require user interaction (hence, post-processing of DCE-MRI studies cannot be seamlessly executed just after acquiring the scans). In this section, we summarize the state of the art for two critical DCE-MRI processing steps---brain tumor segmentation (Section~\ref{sec:brain_tumor_segmentation_SOTA}), and pharmacokinetic modeling (Section~\ref{sec:pharmacokinetic_modeling_SOTA}).

\subsection{Brain tumor segmentation}\label{sec:brain_tumor_segmentation_SOTA}

Approaches for brain tumor segmentation can be divided into four categories---\emph{atlas-based}, \emph{unsupervised}, \emph{supervised}, and \emph{hybrid}. In the \emph{atlas-based} approaches, manually segmented images (\emph{atlases}) are exploited to segment unseen scans~\cite{Pipitone2014494}. These atlases model the anatomical variability of the brain tissue~\cite{Park2014217}. Atlas images are extrapolated to new frames by warping and applying various non-rigid registration techniques. A shortcoming of such techniques is the necessity of creating large (and representative) annotated sets. It is very time-consuming and may lead to atlases which cannot be applied to other tumors~\cite{Aljabar2009726}. Atlas-based approaches are thus most often used for processing images of healthy tissue~\cite{Bauer2010}.

\emph{Unsupervised} algorithms reveal hidden structures within unlabeled data~\cite{Fan2005}. Self-organizing maps are exploited to separate healthy from pathological brain tissue~\cite{Mei201578}. This method has been shown to be able to highlight not only lesions, but also different tissue types within lesions, e.g.,~edema/peritumoral infiltration or necrosis. However, lesions tend to be over-segmented. In evolutionary algorithms~\cite{Taherdangkoo2013}, image segmentation is seen as an optimization problem, in which pixels (or voxels) of similar characteristics are searched. It is tackled in a biologically-inspired manner, in which a population of candidate solutions (e.g.,~pixel or voxel labels) evolves in time. Such techniques commonly require a practitioner to tune their parameters~\cite{Chander20114998}. Other unsupervised algorithms include clustering-based techniques~\cite{Saha2007,Shiyong2014} (also exploiting superpixels~\cite{Verma2013}), and Gaussian mixture modeling~\cite{Simi20151105}.

\emph{Supervised} techniques utilize manually segmented image sets to train a model. These approaches exploit a variety of well-established engines, including decision forests~\cite{Geremia2011378}, conditional random fields~\cite{Wu2014}, and support vector machines~\cite{Ladgham2013}. In~\cite{Zikic2012}, decision forests benefit from context-aware features. These forests are fed with a generative model of tissue appearance elaborated using Gaussian mixture models---the resulting model is quite lightweight, and the algorithm does not require heavy pre-processing (there is no explicit regularization). Appearance and context-based features have also been also utilized in~\cite{Pinto2015}, in their extremely randomized forests. Various deep networks have been successful in segmentation of different kinds of medical images~\cite{Liskowski,SUN201758}, including brain tissue~\cite{Moeskops2016,DBLP:conf/gecco/LorenzoN18}. Hyperintense regions in FLAIR MRI have been segmented using deep autoencoders in~\cite{Korfiatis2016}, where the authors performed extensive preprocessing (skull stripping, N4 bias correction, and intensity standardization). Holistically nested neural nets for MRI were introduced in~\cite{Zhuge2017}. White matter hyperintensities were also segmented in~\cite{Ghafoorian16} using several CNNs which were fed with multiscale patches. Transfer learning was also applied for lesion segmentation in~\cite{Ghafoorian17}. Finally, the winning BraTS'17 algorithm utilized an ensemble of deep neural nets~\cite{10.1007/978-3-319-75238-9_38}. Unfortunately, the authors reported neither training nor inference times of their method which may easily explode for sufficiently deep architectures, and make it inapplicable in a hospital setting.

\emph{Hybrid} approaches combine methods from other categories, e.g.,~atlas-based and unsupervised~\cite{Soltaninejad2017}. They are often tailored to detect a specific lesion type. Superpixel processing was coupled with support vector machines and extremely randomized trees in~\cite{Soltaninejad2017}. Although the results appeared promising, the authors did not report the computation time nor cost of their method, and did not provide any insights into the classifier parameters. Tuning such kernel-based algorithms is very difficult and expensive in practice~\cite{Nalepa2016113}, and improperly selected parameters can easily deteriorate the classifier abilities~\cite{Nalepa2014EVO}. In \cite{RAJENDRAN2012327}, the authors pointed out that deformable models are extensively used for brain tumor segmentation. However, such methods suffer from poor convergence to lesion boundaries. DNNs were coupled with conditional random fields in~\cite{Xiaomei2017}.

\subsection{Pharmacokinetic modeling}\label{sec:pharmacokinetic_modeling_SOTA}

Modern MR scanners provide images suitable not only for qualitative assessment by a reader (to reveal the structural information about the patient), but they are also fast enough to acquire volumetric brain images in relatively short time intervals for the contrast concentration analysis. High spatial and temporal resolutions give a possibility to quantify the concentration of a contrast agent (CA) in tissues, and to assess its distribution in time in terms of a pharmacokinetic model.

In order to apply any pharmacokinetic model to a series of MR images, we use a mapping between the pixel intensity (in an image), and the contrast concentration in the corresponding volume~\cite{Chao2017}. This procedure exploits the CA's magnetic relaxivity (specific for the used agent), the value of patient's haematocrit (HTC), and the pre-contrast T10 relaxation times of the scanned tissue. The HTC value, if not provided, it is assumed to be 0.45~\cite{Chao2017}, whereas the pre-contrast T10 relaxation times are derived from scans at different flip angles of the magnetic field---at least two sequences acquired at two different angles are required. For more details, see~\cite{Chao2017}.

Once the mapping is established, we can obtain quantitative information about the CA's concentration in any MRI series within the analyzed study, including these taken at different moments in time, revealing the kinetics of the CA. We interpret the spatial and temporal information in terms of the Tofts model~\cite{Tofts1991}, which belongs to a group of the compartments models widely used in the DCE analysis. Two compartments of the model represent blood plasma and abnormal extravascular extracellular space (EES). The model allows us to describe the CA kinetics via three tissue parameters, two of which are independent: 1) the influx volume transfer constant $\Ktrans$, or the permeability surface area product per unit volume of tissue between plasma and EES; 2) the volume of EES per unit volume of tissue $\ve$ ($0 \le \ve \le 1$); and 3) the efflux rate constant $k_{ep} = \Ktrans/\ve$. Those parameters are commonly used as biomarkers in quantification of a state of a tumor~\cite{Abe2015}. The model consists of a plasma volume ($v_p$), which is connected to a large EES, and lesion leakage space (LLS). The LLS is assumed to be small enough to not change the total CA concentration, and is connected to the plasma through a leaky membrane. The whole system is assumed to be interconnected, and the CA is well-mixed with plasma. The CA flows to EES and LLS, and it is constantly being depleted by kidneys~\cite{Tofts1991}. In each moment in time, the CA concentration in LLS ($C_t (t)$) is in a dynamic equilibrium, and can be derived from:

\begin{equation}
C_t (t) = \Ktrans \cdot \big( C_p (t) * \exp(-k_{ep} t) \big),
\label{eq:tofts1}
\end{equation}

\noindent where the $*$ symbol between $C_p (t)$ and the exponential decay denotes a convolution operation.

The CA's concentration $C_t (t)$ in LLS can be calculated in two ways: 1) by a numerical convolution, which is computationally expensive; 2) analytically, by exploiting a $C_p (t)$ model and finding an analytical solution to the convolution operation (Eq.~\ref{eq:tofts1}). We take the latter approach, as it may smoothen out the noise of real-life data~\cite{Orton2008}, and is much faster (hence can be deployed in medical applications).

\subsubsection{Bi-exponential model of a vascular input function (VIF)}

Inaccurate modeling of the VIF propagates through to the estimated tissue parameters~\cite{Port2001}, hence an accurate model is required to obtain medical-grade performance. Tofts and Kermode proposed a bi-exponential model of a VIF~\cite{Tofts1991}:
\begin{equation}
	C_p (t) = A \cdot \exp(-\alpha t) + B \cdot \exp(-\beta t),
	\label{eq:BiexponentialModel}
\end{equation}
which has been widely adopted due to its simplicity. However, it assumes that $C_p (t = 0) = \text{max}(C_p)$, which is unrealistic, but was applicable when the scanners had slow sampling rates. Nowadays, it is not the case, as the scanners produce images with temporal resolution which is high enough to track the initial increase of $C_p$ as the contrast begins to arrive.

\subsubsection{Linear model of a VIF}

Orton \textit{et al.} proposed a more realistic model, containing a linear term in $t$, which was called the ``Model 2'' (bi-exponential model was called ``Model 1'') in~\cite{Orton2008}, but will be referred to as a linear model in this work:
\begin{equation}
	C_p (t) = A \cdot t\exp(-\alpha t) + B \cdot \big( \exp(-\beta t) - \exp(-\alpha t) \big).
	\label{eq:LinearModel}
\end{equation}
The linear model has a desired property of $C_p (t = 0) = 0$, which allows for modeling more realistic VIF functions, while maintaining low number of parameters ($A$, $B$, $\alpha$, and $\beta$). Our experiments on a simulated benchmark dataset created by the Quantitative Imaging Biomarkers Alliance (QIBA dataset; Section~\ref{sec:qiba_dataset}) revealed that the linear model outperforms the bi-exponential one, yet it still does not fit the data perfectly.

Although clinically-adopted software for pharmacokinetic modeling is limited (Tissue4D by Siemens is widely used), there exist other implementations which are being validated against benchmark data. Smith \textit{et al.}~\cite{Smith2015} developed the \texttt{DCE-MRI.jl} software suite, which was shown to be overcoming the drawbacks of other DCE-MRI analysis packages (lack of portability, huge computational burden, and complexity being their most important drawbacks).

\section{Method}\label{sec:method}

In this section, we describe the core components of ECONIB: deep learning-powered brain tumor segmentation, determination of the vascular input region, and pharmacokinetic modeling with the proposed VIF cubic model.

\subsection{Brain tumor segmentation}\label{sec:brain_tumor_segmentation}

\begin{figure}[ht!]
\includegraphics[width=1\columnwidth]{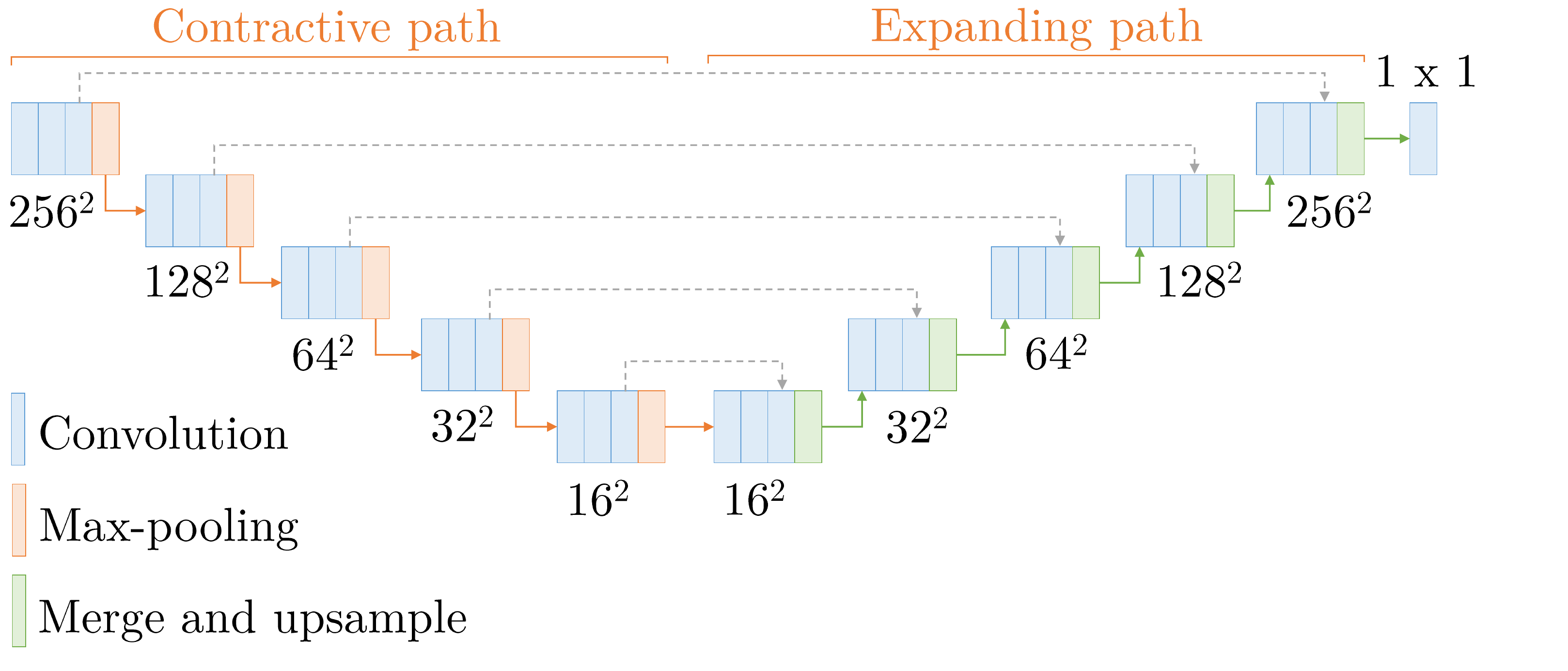}
\caption[fig:unet]{Our U-Net-based DNN with its blocks and connections.}
\label{fig:unet}
\end{figure}

Our DNN designed for brain tumor segmentation, illustrated in Fig.~\ref{fig:unet}, consists of a series of blocks placed symmetrically as a \emph{contractive} and \emph{expanding} path, yielding a U-shape~\cite{RonnebergerFB15}. Each block in the contractive path contains three $3\times 3$ convolution layers with 64 feature maps (filters) each, followed by a rectified linear unit (ReLU) activation, and a max-pooling layer with $2\times 2$ kernel and stride (offset) to perform downsampling. In the expanding path, the output of each block receives a skip connection from the depth-matched feature maps (from the contractive path), and concatenates it to its own output, which is upsampled and passed to a higher block. The last layer of the network is a $1\times 1$ convolutional layer with sigmoid activation, which reduces the activation depth to one.

In this manner, high-level features, extracted in the contractive path, propagate through higher resolution layers of the expanding path. In Fig.~\ref{fig:unet}, we can observe how the size of the feature map is affected by each operation, achieving a compression of the feature space from $256\times 256$ to $16\times 16$. Residual connections between corresponding blocks in the contractive and expanding paths are also displayed. 
Our DNN performs multi-scale analysis---features from the contractive path are combined with the upsampled output, where a large number of available channels allows for context propagation through high-resolution layers. As a consequence, the expanding path is symmetric to the contractive path. The total number of trainable parameters of our network is close to $9\cdot 10^5$. Interestingly, our model can be applied to multi-modal processing (e.g.,~more than one image modalities can be used for training and inference by stacking them altogether in a multi-channel image)---in this work, we experimentally prove this flexibility by segmenting brain tumors from MRI scans with three deep models trained using one, three, and four modalities (see details in Section~\ref{sec:validation_brain}).

Our model is inspired by the U-Net architecture which has been already applied for biomedical image processing~\cite{RonnebergerFB15}. There are two main differences between the proposed DNN and the state-of-the-art U-Net---first, the number of filters is constant at each step of the processing pipeline in our model, while it is doubled in each deeper block in the original U-Net. Keeping a constant number of filters reduces the number of parameters of the network, effectively lowering its computational requirements and processing time. Second, we preserve the shape of each feature map, which allows us to seamlessly take advantage of the bridged connections by simply concatenating activation maps at each depth (cropping is employed in~\cite{RonnebergerFB15}).

\subsection{Determination of vascular input region}\label{sec:sagittal_sinus_segmentation}

\begin{figure}[ht!]
\centering
\includegraphics[width=0.7\columnwidth]{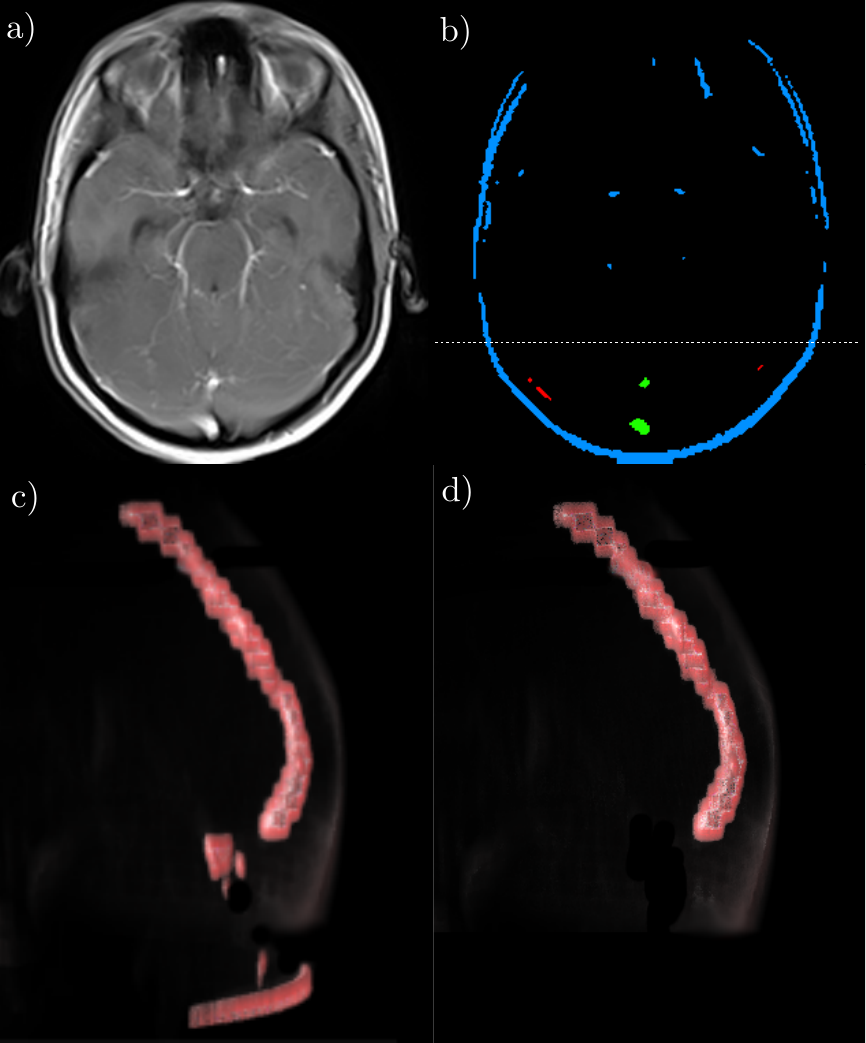}
\caption[fig:unet]{Determination of the vascular input region: a)~a slice in the axial plane from an input volume, b)~intensity thresholding reveals candidate regions, from which: blue regions are rejected because they occupy an upper section of the image (above the dotted line), red ones are rejected due to the shape irregularities, green regions are retained, c)~visualization of the volume with remained voxels grouped as connected components, d)~visualization of the volume after retaining only the largest 3D connected component.}
\label{fig:sss}
\end{figure}

The input of our algorithm is a time series of co-registered T1 VIBE scans. First, an arithmetic average $\mu_{I}$ is calculated for high-intensity voxels for each volume. Voxels are considered to be high-intensity, if their intensity is above a given threshold (we manually tuned this parameter to $0.75$ of the maximum intensity in a volume). A volume with the maximum $\mu_{I}$ is selected from the time series as $V_{s}$.
Next, $V_{s}$ is cropped (Fig.~\ref{fig:sss}a), and a binary mask with the high-intensity voxels is created for $V_{s}$, yielding the volume $V_{T}$.
For each slice from $V_{T}$, we perform blob detection. To narrow down the search space for the vascular input region, only the components in the lower section of the slice are retained (Fig.~\ref{fig:sss}b), which corresponds to the $G$, $H$, and $I$ sections of the brain in Talairach coordinates. We introduce a simple shape metric: $M(S) = A_{c}/A_{b}$, where $A_{c}$ is the area of a shape $S$, and $A_{b}$ is the area of its bounding box. For a square shape, the metric becomes 1, whereas for a circle it is $\pi/4$ (for elongated and curvilinear shapes, the value of the metric will be lower). All connected components $S_i$ for which $M(S_i)<\pi/4$ are rejected (Fig.~\ref{fig:sss}c). Finally, the binary volume $V_{T}$ undergoes the 3D connected-components labeling, and the component with the largest volume is considered to contain only the voxels of the vascular region of interest in $V_{s}$ (Fig.~\ref{fig:sss}d). We propagate the binary labels from $V_{T}$ to all volumes (they are co-registered), and use them to measure the contrast concentration. Importantly, our algorithm is deterministic and delivers reproducible vascular input region determination.

\subsection{Pharmacokinetic modeling}\label{sec:pharmacokinetic_modeling}


In this work, we propose a generalization of the linear model (with the aim of minimizing the fitting error) by substituting $t \rightarrow t^n$ mentioned in~\cite{Orton2008}, and putting $n = 3$. We have also investigated a model with a $t^2$ term, however it yielded more inaccurate results. We call the model with the $t^3$ term the cubic model. To the best of our knowledge, such model has not been used before. It also has only four parameters ($A, B, \alpha$, and $\beta$):

\begin{equation}
C_p (t) = A\cdot t^3 \exp(-\alpha t) + B \cdot\big( \exp(-\beta t) - \exp(-\alpha t) \big).
\label{eq:CubicModelPlasma}
\end{equation}

\noindent It preserves the property of $C_p (t = 0) = 0$, while its form allows for finding an analytical solution to Eq.~\ref{eq:tofts1}. $C_t (t)$ is parametrized by six parameters: $A, B, \alpha, \beta$ originating from a VIF, and $\Ktrans$ and $k_{ep}$ (or $\ve = \Ktrans/k_{ep}$):

\begin{gather}
C_t = \Ktrans \big( A \cdot \Delta^{-4}\exp(-\alpha t) \cdot C_{t1} + B \cdot C_{t2} \big), \nonumber \\
C_{t1} = \big( -(t \Delta)^3 - 3(t \Delta)^2 - 6 t \Delta - 6 \big) \exp (\Delta) + 6, \nonumber \\
C_{t2} = \frac{\exp (-\beta t) - \exp (-k_{ep} t)}{k_{ep} - \beta}\nonumber \\ - \frac{\exp (-\alpha t) - \exp (- k_{ep}t)}{k_{ep} - \alpha},
\label{eq:CubicModelTissue}
\end{gather}
\noindent where $\Delta = \alpha - k_{ep}$.

\begin{figure}[ht!]
\centering
	\includegraphics[width=0.7\columnwidth]{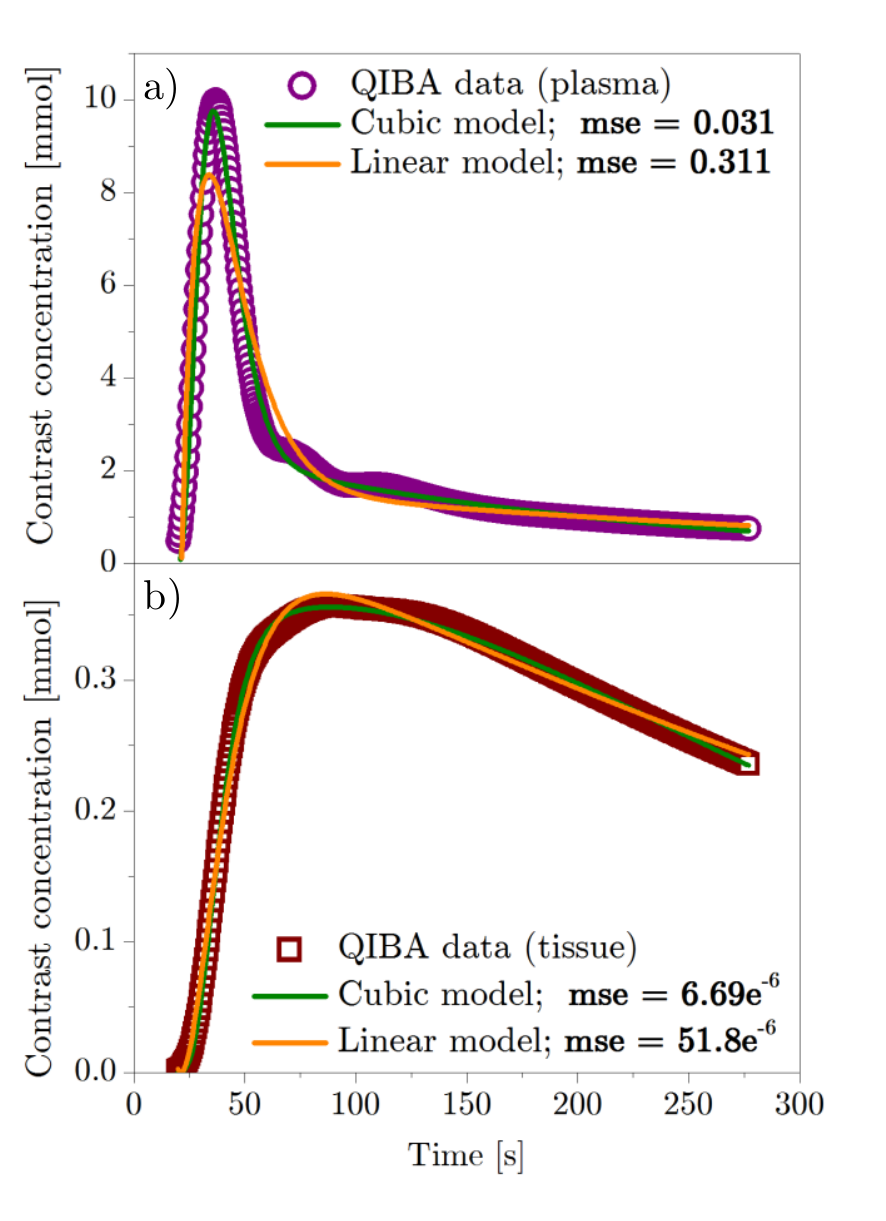}
	\caption[fig:AIFmodels]{Fits of the linear model and our cubic model of the VIF to the QIBA phantom data: contrast agent's concentration in a) vascular input region, and in b) tissue. The data represents tissue characterized by $\Ktrans = 0.10$ $\text{min}^{-1}$ and $\ve = 0.20$. The mean square error (mse) of the fit is written in bold.}
	\label{fig:AIFmodels}
\end{figure}

The comparison of fits of the linear and cubic models to the QIBA data is presented in Fig.~\ref{fig:AIFmodels}. The function of the contrast concentration in time in the vascular input region and tissue with the fitted curves is presented on panels a) and b), respectively. The cubic model (green curve) has around an order of magnitude lower mean square error (mse) than the fit of the linear model (orange curve), for both plasma and tissue. This empirical evidence shows that our proposed model of a VIF has a potential to yield higher-quality fits required to obtain tissue parameters with high precision. The evaluation of the cubic model on the QIBA dataset is presented in detail in Section~\ref{sec:benchmark_datasets_qiba}.

%

\section{Data}\label{sec:datasets}

\subsection{QIBA phantom dataset}\label{sec:qiba_dataset}

The Quantitative Imaging Biomarkers Alliance published virtual DCE phantom sets that can be used for validating DCE-MRI analysis approaches\footnote{\url{https://qibawiki.rsna.org/index.php/Synthetic_DCE-MRI_Data}}. Several phantoms are available to validate both the procedure of mapping MR pixel intensities to the contrast concentration values, and tissue parameters fitting. In this work, we exploit the newest release of the QIBA set (version 14) simulating images obtained for  patients who have low cardiac output, hence the patients would be expected to have lower and ``broader'' input functions. This set consists of 661 DICOM files (each file simulating data of one timestamp). The data in each DICOM is divided into two regions---tissue and vascular. The tissue part is further divided into 30 (5$\times$6) non-overlapping patches of size $10 \times 10$ pixels, marking regions of tissue characterized by different values of $\Ktrans \in \{0.01, 0.02, 0.05, 0.10, 0.20, 0.35\}$ $\text{min}^{-1}$, and $\ve \in \{0.01, 0.05, 0.10, 0.20, 0.50\}$. The vascular part is placed at the bottom of an image, and occupies the area of $10 \times 50$ pixels (Fig.~\ref{fig:QIBAbenchmark_results}a). Examples of CA concentrations as a function of time in plasma and tissue for one patch ($\Ktrans = 0.10$ $\text{min}^{-1}$ and $\ve = 0.20$) are shown in Fig.~\ref{fig:AIFmodels}a and Fig.~\ref{fig:AIFmodels}b.

QIBA does not provide the reference values for the vascular region curve (the authors do not recommend using any particular model for VIF, nor any method to process the data). The only way to verify whether a given VIF model yields desired performance is to test the complete solution and compare the resulting values of the tissue parameters with those provided by QIBA---we follow this approach in this work.

\subsection{Brain tumor segmentation benchmark dataset (BraTS'17)}

The performance of our DNN segmentation was evaluated over the newest release of the famous Brain Tumor Segmentation (BraTS) dataset~\cite{Menze2015,Bakas2017} (version 2017). The BraTS'17 set contains DCE-MRI data of 285 patients with diagnosed gliomas---210 high-grade glioblastomas (HGG), and 75 low-grade gliomas (LGG). The data comes in four co-registered modalities---native pre-contrast (T1), post-contrast T1-weighted (T1c), T2-weighted (T2), and T2 Fluid Attenuated Inversion Recovery (FLAIR). All the pixels have one (out of four) label: healthy tissue, Gd-enhancing tumor (ET), peritumoral edema (ED), the necrotic and non-enhancing tumor core (NCR/NET). As we are interested only in a binary classification (healthy/non-healthy), the classes ET, ED, and NCR/NET are merged together into the ``non-healthy'' class.

\subsection{Clinical DCE-MRI data}

ECONIB was validated over the data collected for patients who underwent the MR imaging with a MAGNETOM Prisma 3T system (Siemens, Erlangen, Germany) equipped with a maximum field gradient strength of 80 mT/m, and using a 20-channel quadrature head coil. The MRI sequences were acquired in the axial plane with a field of view of $230\times190$ mm, matrix size $256\times256$ and 1 mm slice thickness with no slice gap. In particular, we exploited the series with TE=386 ms, TR=5000 ms, and inversion time of 1800 ms for segmentation of brain tumors. 44 consecutive BT LGG WHO II patients (age: 40.8$\pm$13.9 years, 23 males) were analyzed retrospectively by an experienced reader (8 years of experience) who provided us with the ground-truth masks used for validating our brain tumor segmentation.

\section{Experimental validation}\label{sec:experimental_validation}

\subsection{Experimental setup}\label{sec:experimental_setup}

We validated the pivotal ECONIB components separately over benchmark and clinical sets to check their robustness against different data (see Sections~\ref{sec:validation_brain}--\ref{sec:benchmark_datasets_qiba}), and presented an example of a full end-to-end processing in Section~\ref{sec:example_dce_MRI_full}.

ECONIB was coded in \texttt{C++} and compiled with Microsoft Visual C++ 2017 (for segmentation, we used \texttt{Tensorflow} 1.7 over CUDA 9.0 and CuDNN 5.1). The experiments ran on an Intel i7-6850K (15 MB Cache, 3.80 GHz) CPU machine with 32 GB RAM and NVIDIA GTX Titan X GPU with 12 GB VRAM. We trained our DNN using Nadam~\cite{Dozat2015} (the initial learning rate was $10^{-5}$, and optimizer parameters were $\beta_1 = 0.9$, $\beta_2 = 0.999$). The training ran until DICE (Section~\ref{sec:brats_experiments}) over the validation set did not increase by at least 0.002 in 10 consecutive epochs. For segmentation, we binarize the DNN predictions (threshold of 0.5), and process them using the connected-components analysis to get rid of small false-positive regions.

\subsection{Validation of brain tumor segmentation}\label{sec:validation_brain}

\subsubsection{BraTS'17 dataset}\label{sec:brats_experiments}

In this experiment, we validated our DNN for brain tumor segmentation over the BraTS'17 dataset. First, we performed a 6-fold cross-validation by splitting the dataset into 205-40-40 (training-validation-testing) volumes (patients). We exploited images of three modalities (FLAIR, T1c, and T2-weighted), stacked together as channels of a single image. To evaluate the segmentation performance, we use the DICE score which is calculated as
  ${\rm DICE(A,B)}=\frac{2 \cdot \left|A\cap B\right|}{\left|A\right|+\left|B\right|}$,
where $A$ and $B$ are two segmentations, i.e.,~manual and automated. DICE ranges from zero to one (one is the perfect score). We report \emph{precision} $\Precision$ (percentage of correctly classified pixels out of all the pixels classified as lesions), and \emph{sensitivity} $\Recall$ (percentage of lesion pixels correctly classified as lesions) elaborated by the models. They are given as $\Precision={\rm \frac{TP}{TP+FP}}$ and $\Recall={\rm \frac{TP}{FN+TP}}$, where TP, FP, and FN are the numbers of true positives, false positives, and false negatives, respectively.

The results are gathered in Table~\ref{tab:BRATScrossvalidation}---the average DICE score (over the unseen testing set) is 0.8293, and its standard deviation amounts to 0.0252 (hence, the segmentation performance of a trained model is very high for all folds). Consequently, both sensitivity and precision are well above 0.8, with the best fold surpassing 0.93 in precision (with sensitivity exceeding 0.82). It shows that our deep model delivers accurate and fairly consistent segmentation, and can be robustly trained using different MRI data (the folds were non-overlapping).

\begin{table}[ht!]
	\centering
	\caption{The segmentation performance of our DNN over the BraTS'17 test set (in a 6-fold cross-validation setting). The best and the worst folds are selected with respect to the DICE score (the higher the score is, the better).}
	\label{tab:BRATScrossvalidation}
\renewcommand{\tabcolsep}{5.5mm}
	\begin{tabular}{llcc}
		\Xhline{2\arrayrulewidth}
		& DICE     & Sensitivity  &  Precision  \\
		\hline
		Mean       & 0.8293   & 0.8102 & 0.8842 \\
		Std. Dev.  & 0.0252   & 0.0464 & 0.0346 \\
		\hline
		Best fold  & 0.8664   & 0.8232 & 0.9323 \\
		Worst fold & 0.7798   & 0.7373 & 0.8789 \\		
		\Xhline{2\arrayrulewidth}
	\end{tabular}
\end{table}

To verify the generalization capabilities of our model, we segmented the validation set, containing data of 46 patients without the ground-truth segmentation\footnote{Our segmentations were automatically assessed by the BraTS organizers. Note that the organizers do not report precision for the validation set.}. Additionally, we verified whether adding one additional modality (T1) can improve the overall performance, therefore we trained two models of the same architecture, accepting images with 3 and 4 channels. In both cases, the dataset was split into the training and validation sets, containing 245 and 40 volumes, respectively. Before feeding into the DNN, the images were z-normalized (with the mean set to 0, and the standard deviation to 1). The details of the training process are presented in Table~\ref{tab:BRATStraining}. The number of epochs required to converge differ, which is attributed to the stopping condition and random initialization of the network's weights. However, the processing time of one epoch is similar (adding one modality influences the number of parameters only in the first layer, which is not significant with respect to the total number of the network parameters).

\begin{table}[ht!]
	\centering
	\caption{The number of training epochs until reaching convergence, time required to process a single epoch, and to train our DNN over BraTS'17 using 3 and 4 modalities.}
	\label{tab:BRATStraining}
\renewcommand{\tabcolsep}{4mm}
	\begin{tabular}{cccc}
		\Xhline{2\arrayrulewidth}
		Modalities & Epochs & One epoch & Total \\
		\hline
		FLAIR, T1c, T2      & 32     & 791 s  & 7 h 02 m  \\
		FLAIR, T1, T1c, T2  & 24     & 798 s  & 5 h 19 m   \\
		\Xhline{2\arrayrulewidth}
	\end{tabular}
\end{table}

The results obtained for both DNN versions (with 3 and 4 modalities) are gathered in Table~\ref{tab:BRATSvalidation} (for the unseen data, the BraTS organizers report specificity as well, which is given as $\Specificity={\rm \frac{TN}{TN+FP}}$). The mean DICE score (over 46 patients) is close to 0.83 in both cases, which is also consistent with the mean score obtained during cross-validation. The results indicate that there is no statistical difference between using three or four modalities (Wilcoxon test, $p<0.01$). Although the best-performing state-of-the-art segmentation method delivers slightly better DICE (0.901)~\cite{10.1007/978-3-319-75238-9_38}, the authors reported neither training nor inference times of their technique (since it was an ensemble of deep networks, it may be inapplicable in practice). Interestingly, our algorithm not only outperforms other techniques from the literature in terms of segmentation accuracy, but is also much faster to learn (e.g.,~3D CNNs obtained 0.822 DICE on average and took 120 h to train~\cite{10.1007/978-3-319-75238-9_20}, whereas conditional adversarial nets obtained 0.7 DICE and required 72 h to train using parallel GPUs~\cite{10.1007/978-3-319-75238-9_21}).

\begin{table}[ht!]
	\centering
	\caption{The segmentation performance over 46 BraTS'17 validation patients obtained using our DNN trained with (a) T1c, T2, and FLAIR, and with (b) T1, T1c, T2, FLAIR.}
	\label{tab:BRATSvalidation}
\renewcommand{\tabcolsep}{4mm}
	\begin{tabular}{llccc}
		\Xhline{2\arrayrulewidth}
        & & DICE     & Sensitivity  & Specificity  \\
		\hline
		    & Mean      & 0.8301   & 0.8391       & 0.9914       \\
		(a) & Std. Dev. & 0.1176   & 0.1390       & 0.0068       \\
		    & Median    & 0.8676   & 0.8752       & 0.9931       \\
		\hline
		& Mean      & 0.8279   & 0.8379       & 0.9917       \\
		(b) & Std. Dev. & 0.1159   & 0.1415       & 0.0065       \\
		& Median    & 0.08657  & 0.8789       & 0.9930      \\
		\Xhline{2\arrayrulewidth}
	\end{tabular}
\end{table}

\begin{figure}[H]
\centering
	\includegraphics[width=0.7\columnwidth]{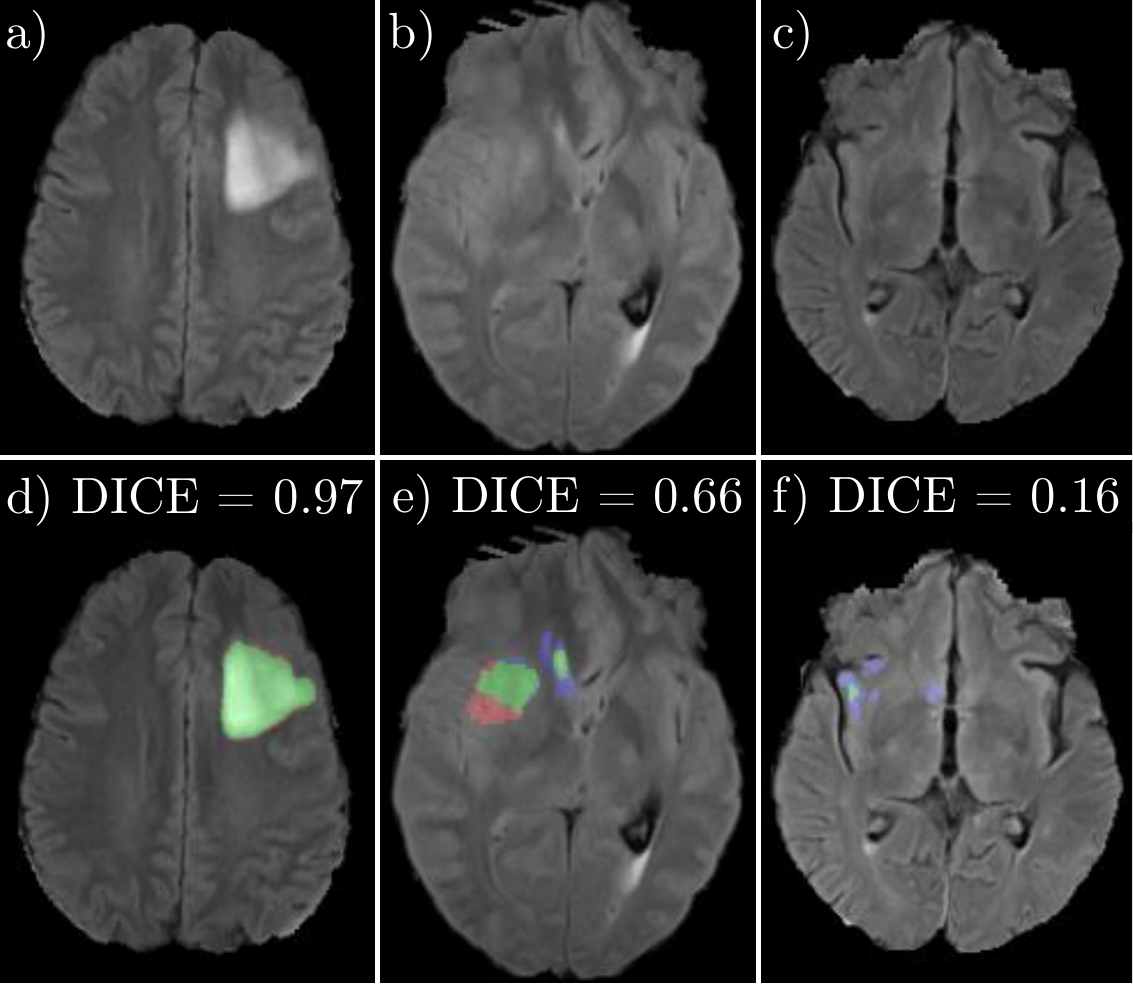}
	\caption[fig:BRATSsegmentation]{Examples of the BraTS'17 images segmented using our DNN: a), b), and c) are original T2-weighted images, d), e), and f) are corresponding segmentations. Green color represents true positives, blue---false negatives, and red---false positives.}
	\label{fig:BRATSsegmentation}
\end{figure}

Examples of segmented (using our model trained with three modalities) BraTS'17 images are rendered in Fig.~\ref{fig:BRATSsegmentation}. The top row (panels a, b, and c) shows original T2-weighted images, whereas the bottom row (panels d, e, and f) visualize segmentations overlaid over the original images with colors indicating the true positives (in green), false positives (red), and false negatives (blue). For larger tumors, the quality of our segmentation is usually higher (reaching as high as 0.97 DICE), while for smaller and less pronounced lesions our DNN tends to under-segment. Although the DICE score is very low in Fig.~\ref{fig:BRATSsegmentation}f (only 0.16), the tumor is appropriately detected and could be further analyzed by a human reader (e.g., for improving segmentation if necessary). The time required to process one volume (155 images, $240 \times 240$ pixels each) from this set is 68 s on a CPU, and 7 s on a GPU.

\subsubsection{Clinical DCE-MRI data}\label{sec:clinical_dce_data}

In this experiment, we validated our DNN using a clinical set of 44 LGG patients. The original scans, provided in various resolutions, were reshaped to $256 \times 256$ preserving the aspect ratio (symmetric zero-padding was added if necessary). In order to assess the performance of our DNN, the data was split into the training, validation, and testing sets containing 28, 8, and 8 patients (without the overlaps), and a 6-fold cross-validation was performed. The training (using only T2 sequences) took $33\pm 6$ epochs, which corresponds to $75\pm 15$ minutes on average per fold while exploiting a single GPU.

\begin{table}[ht!]
	\centering
	\caption{The performance of our DNN over the clinical data (in a 6-fold cross-validation setting). The best and the worst folds are selected with respect to the DICE score.}
	\label{tab:WHOvalidation}
\renewcommand{\tabcolsep}{2mm}
	\begin{tabular}{llcccc}
		\Xhline{2\arrayrulewidth}
		& DICE     & Sensitivity  & Specificity & Precision & FN rate \\
		\hline
		Mean       & 0.7552   & 0.7111	& 0.9996	& 0.8626 & 0.1401 \\
		Std. Dev.  & 0.0727   & 0.1164 & 0.0001    & 0.0502  & 0.0536 \\
		\hline
		Best fold  & 0.8464   & 0.8685 & 0.9996    & 0.8296 & 0.1125 \\
		Worst fold & 0.6254   & 0.5191 & 0.9997    & 0.9224 & 0.1779 \\		
		\Xhline{2\arrayrulewidth}
	\end{tabular}
\end{table}

\begin{figure}[ht!]
\centering
	\includegraphics[width=0.7\columnwidth]{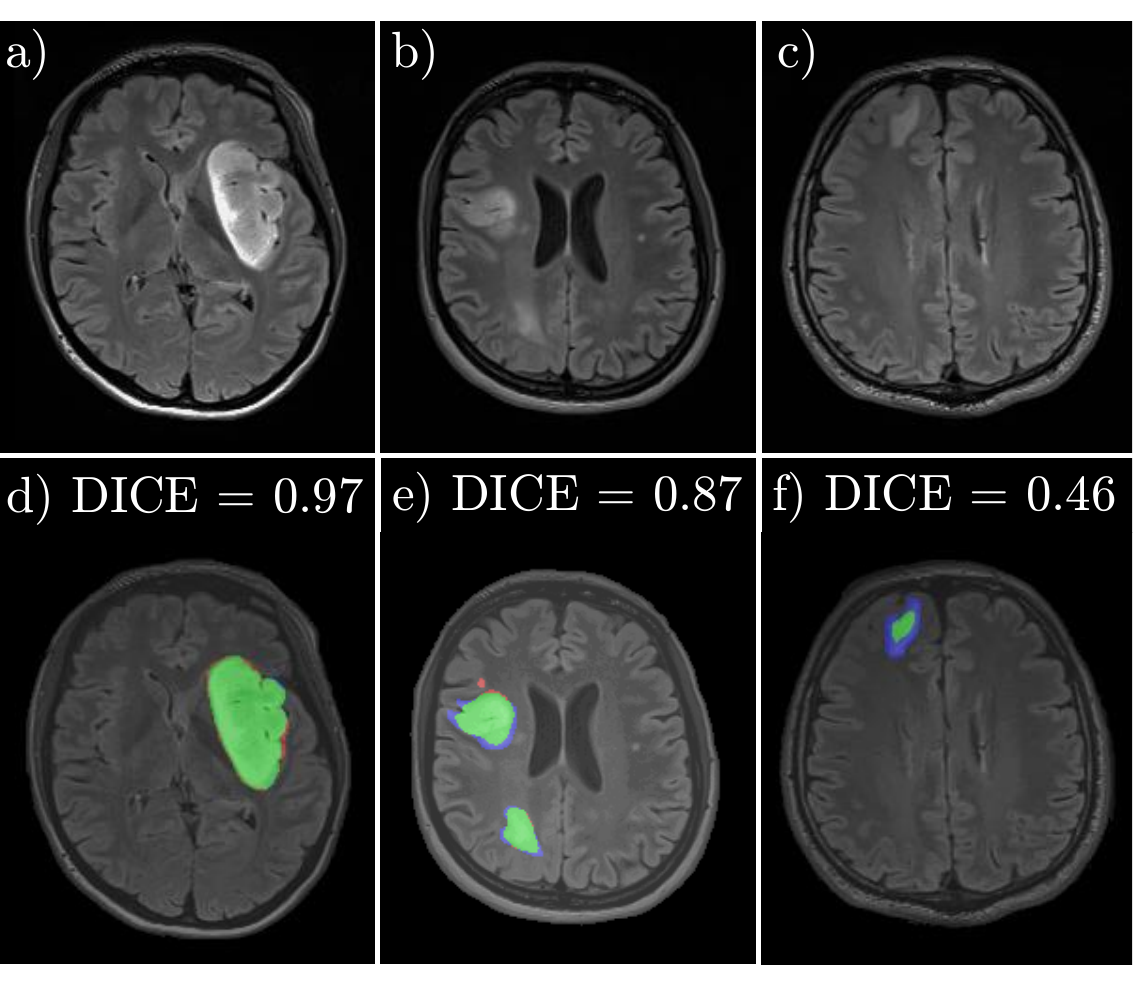}
	\caption[fig:WHOsegmentation]{Examples from our DCE-MRI set segmented using our DNN: a), b), and c) are original T2-weighted images, d), e), and f) are corresponding segmentations. Green color represents true positives, blue---false negatives, and red---false positives.}
	\label{fig:WHOsegmentation}
\end{figure}

\begin{figure*}[ht!]
\centering
\hspace*{-1cm}
\setlength{\tabcolsep}{2pt}
\begin{tabular}{ccc}
a) & b) \\
\includegraphics[width=0.5\columnwidth]{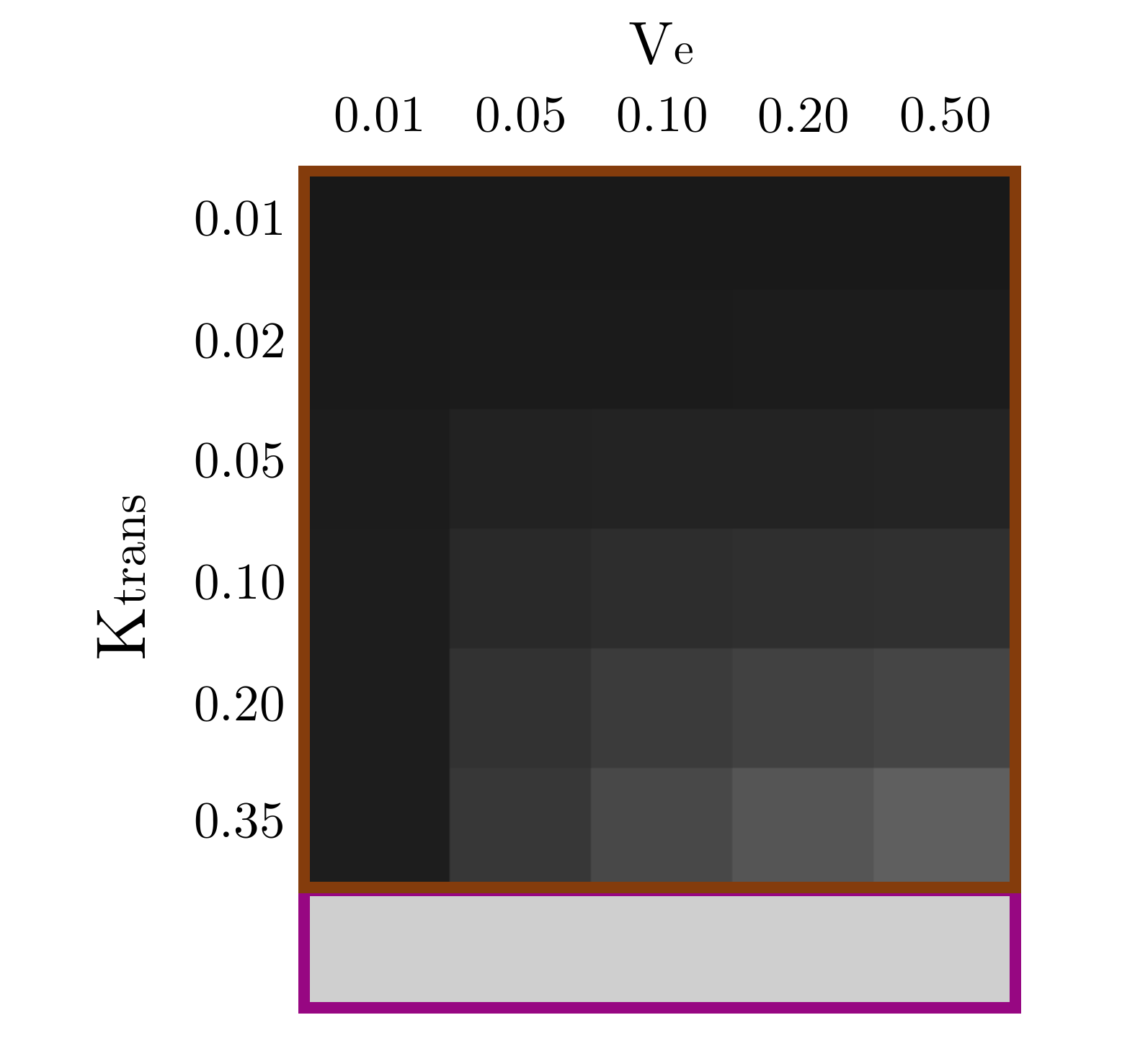} & \includegraphics[width=0.5\columnwidth]{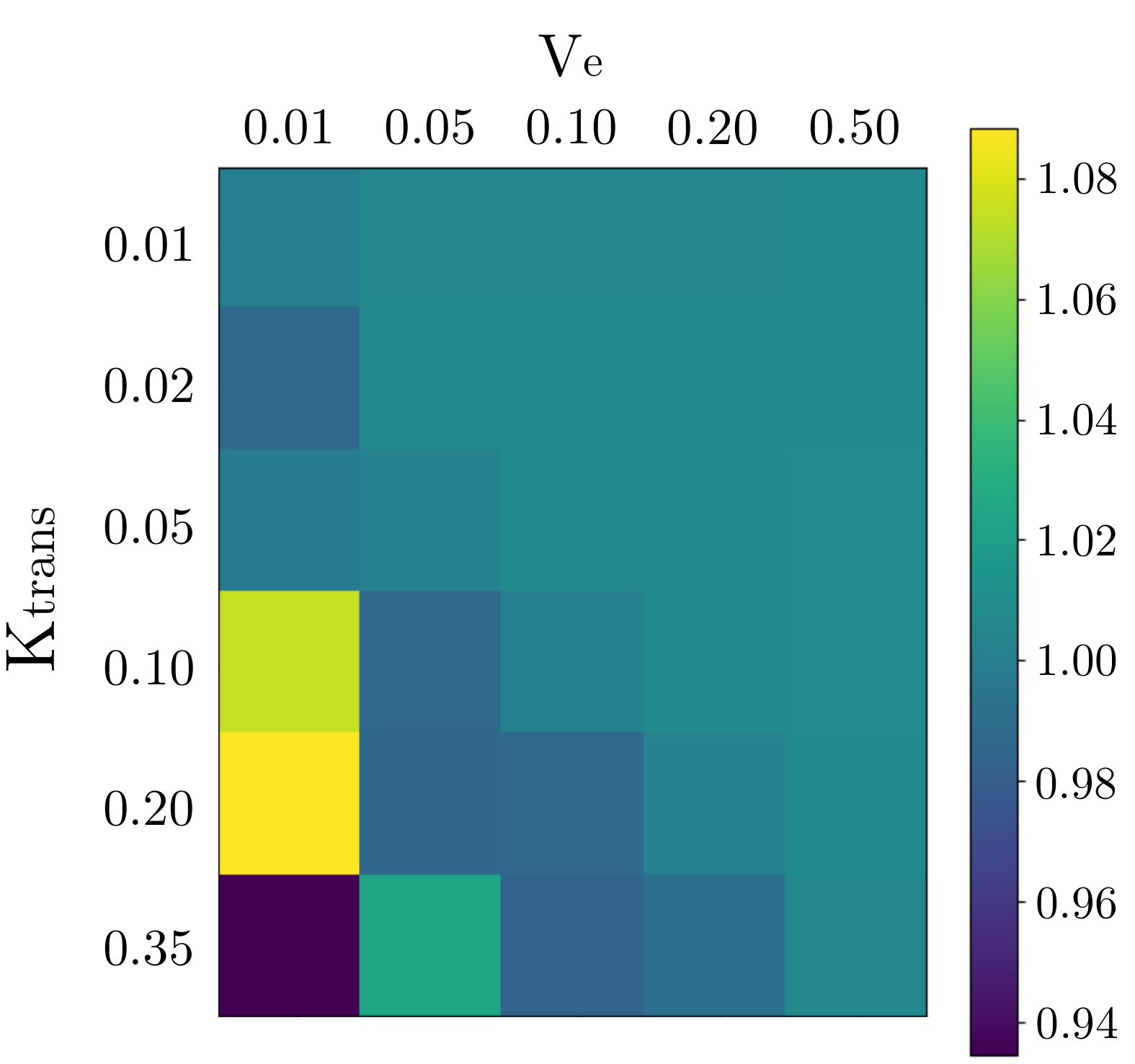} \\
\multicolumn{2}{c}{c)}\\\multicolumn{2}{c}{\includegraphics[width=0.5\columnwidth]{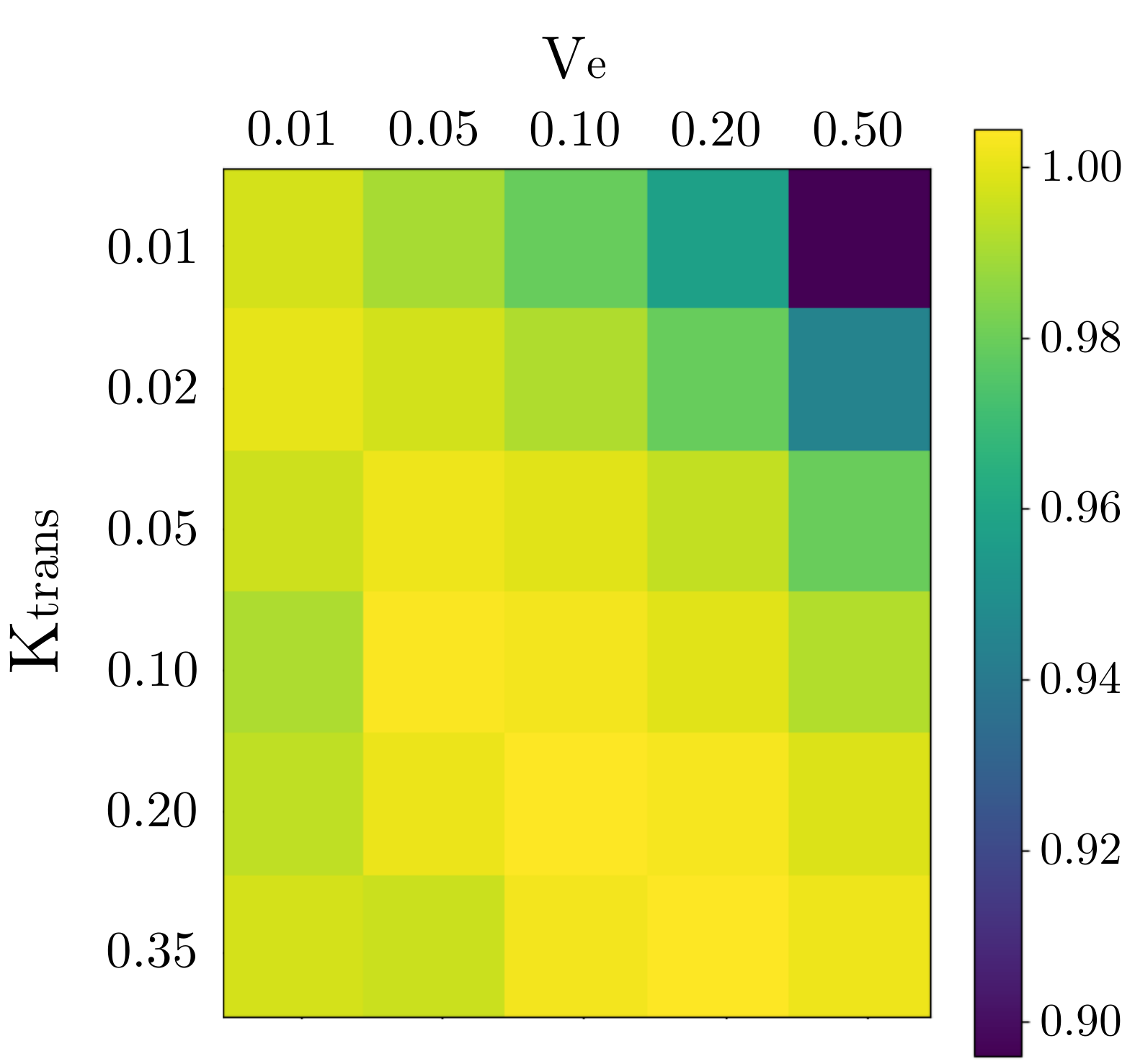}}\\
\end{tabular}
\caption[fig:QIBAbenchmarkResults]{QIBA (version 14) phantom set: a) a visualization of a single DICOM file (brown frame surrounds 10 $\times$ 10 px patches of tissue characterized by a combination of different $\Ktrans$ and $\ve$, purple frame surrounds a $10 \times 50$ px vascular region) alongside the results of fitting the cubic model to this set divided by the expected values of b) $\Ktrans$ and c) $\ve$.}
\label{fig:QIBAbenchmark_results}
\end{figure*}

The results (obtained for the unseen test set) are gathered in Table~\ref{tab:WHOvalidation}. The average DICE score amounts to 0.7552, which is slightly lower than the DICE score obtained over the BraTS'17 dataset. We attribute this difference to two factors---first, our clinical dataset contains $6$ times less patients (hence, the heterogeneity of the training set is significantly lower). Second, the segmentation labels provided by a reader may be biased, which may affect the training and give lower generalization performance (it could be addressed by labeling a training set by multiple readers and agreeing on the final ``gold-standard'' segmentation). Nevertheless, our DNN delivers consistent segmentation across folds (standard deviation is fairly low for all measures) which are relatively small in size, and do not expose high tumor shape and size variability (the worst fold with 0.6254 DICE did not encompass very small tumors which appeared in the test set). In Table~\ref{tab:WHOvalidation}, we also included the FN rate (percentage of frames erroneously annotated as ``healthy'' by our DNN) which shows that ECONIB agrees with a reader in 86\% of all manually-selected tumorous frames on average.

Examples of the clinical DCE-MRI images segmented using our technique are shown in Fig.~\ref{fig:WHOsegmentation}. Similarly, the best segmentation performance is observed for large brain tumors, while smaller ones tend to be under-segmented (due to a small number of such examples in the entire dataset). In Fig.~\ref{fig:WHOsegmentation}b, we can observe that our deep model correctly detects and segments multiple tumors in a single frame as well. The time required to process one volume (160 images) from this set was 77 s on a CPU, and 7 s using a single GPU on average.

\subsection{Validation of the vascular input region determination}\label{sec:validation_sss}
Determination of the vascular input region is an important step in extracting the plasma curve. To verify our algorithm for this task, we calculate the root mean square (RMS) intensity within the VIF regions for each slice separately, and compare the values with the VIF masks segmented by a reader.
%
%
Here, we report the results obtained over our clinical data of 44 patients (see Section~\ref{sec:clinical_dce_data}; 1995 slices contained the segmented vascular input regions in this dataset). The average RMS difference was $2.69 \cdot 10^{-5}$ with the standard deviation amounting to $1.08 \cdot 10^{-2}$, median equal to $-9.40 \cdot 10^{-5}$, and maximum of $0.0481$. The average analysis (of the entire study) takes $7.52\pm2.24$ ms.

\subsection{Validation of pharmacokinetic modeling}\label{sec:benchmark_datasets_qiba}

To quantify the performance of our pharmacokinetic modeling, we calculated the tissue parameters for the QIBA phantom data (for each patch), and divided them by their expected values. The results are presented in Fig.~\ref{fig:QIBAbenchmark_results}(b,c), for $\Ktrans$ and $\ve$, respectively. Our fitting accuracy is very high for most of the simulated regions (the average error given as $\Ktrans_{error}=\left|\Ktrans_{fitted} - \Ktrans_{target}\right|/\Ktrans_{target}$ equals $1.510\%$)---the largest error occurred for the regions with expected high values of $\Ktrans$, and low values of $\ve$---the discrepancy reaches around 6-8\% (Fig.~\ref{fig:QIBAbenchmark_results}b). Analogously, the largest error (approx. 6-8\%) for $\ve$ was visible in the regions of expected high values of $\ve$, and low values of $\Ktrans$ (Fig.~\ref{fig:QIBAbenchmark_results}c). The corresponding average error of the $v_{e}$ fitting was $2.782\%$. The time of fitting 3000 voxels of this QIBA set (ver. 14) was close to 5 s. A corresponding fitting error, using the linear model (Eq.~\ref{eq:LinearModel}), was 6.501\% and 5.261\% for the $\Ktrans$ and $\ve$, respectively.

We compare the efficacy of our modeling with \texttt{DCE-MRI.jl}, whose authors operated on version 6 of the QIBA dataset---their fitting errors over this set were 0.419\% and 0.126\% for $\Ktrans$ and $\ve$, respectively. Importantly, \texttt{DCE-MRI.jl} exploits numerical integration instead of an analytical model to approximate the VIF, which eliminated the approximation error. However, numerical calculations are more computationally demanding than analytical approaches. Despite the fact that the authors implemented their algorithms in Julia, which is claimed to be a high-level language designed for scientific computing~\cite{Smith2015}, the reported time required to fit 3000 voxels of the QIBA set was approx. 520 s using 8 threads. The fitting error of the cubic (linear) model obtained on the QIBA6 dataset was 5.114\% (4.684\%) and 3.241\% (3.466\%) for the $\Ktrans$ and $\ve$, respectively. Our system, while characterized by a worse accuracy, requires less than 5 s to process the QIBA set (version 6), hence it is more than $100\times$ faster than \texttt{DCE-MRI.jl}. Finally, our system is also portable and can be easily deployed using any hardware and operating system.


\begin{figure*}[ht!]
\hspace*{-0.5cm}	\includegraphics[width=1.05\textwidth]{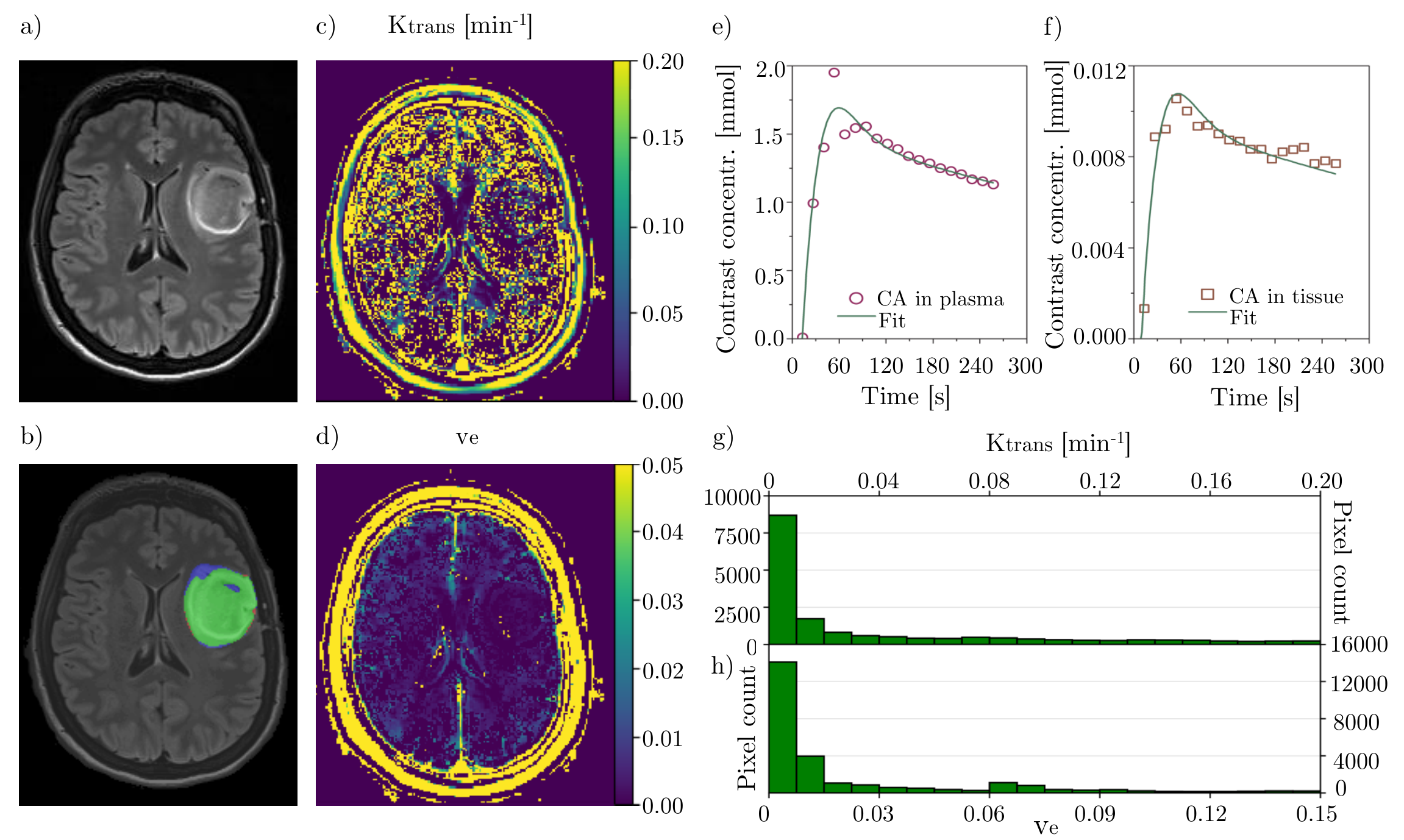}
	\caption[fig:DCEsteps]{Artifacts generated at the pivotal steps of ECONIB: a) T2-weighted image, b) segmented tumor (the meaning of the colors as before), the parameter maps for c) $\Ktrans$, and d) $\ve$, contrast agent's concentration alongside our fitting in e) plasma, and f) tissue, and the histograms extracted from the g) $\Ktrans$ (skewness is 15.859, kurtosis is 258.021), and h) $\ve$ (skewness is 3.571, kurtosis is -15.747) parameter maps. Automated analysis of this patient took less than 3 min with one GPU.}
	\label{fig:DCEsteps}
\end{figure*}

\subsection{Illustrative example of automated DCE-MRI processing}\label{sec:example_dce_MRI_full}

In this section, we present an end-to-end processing example of an LGG patient, alongside execution times of all pivotal steps of ECONIB. In Fig.~\ref{fig:DCEsteps}, we render the artifacts generated at each step of the ECONIB pipeline (the analysis took less than 3 min using a single GPU, and 4 min without GPU). All of them are finally embedded into a DICOM report which can be either sent back to PACS, or analyzed locally by a reader (the process can be triggered automatically from PACS, and does not require changing the clinical protocol). In this example, we showed that ECONIB can extract quantifiable (histogram-based~\cite{Miles2013CI}) features from parameter maps: kurtosis (measure of the peakedness of the histogram) and skewness (measure of the asymmetry of the histogram). However, to use them as biomarkers (e.g.,~in conjunction with $\Ktrans$ and $\ve$), a validation process is required (it is currently ongoing).

\begin{table}[ht!]
	\centering
	\caption{Time required to accomplish selected parts of ECONIB using a single GPU and without GPU (the CPU version). This study consists of 160 T2 images for brain tumor segmentation, and 25 volumes (of 30 T1 images, 192 $\times$ 192 px each) acquired at different timestamps.}
	\label{tab:DCEtimeProfile}
\renewcommand{\tabcolsep}{3mm}
	\begin{tabular}{rcc}
		\Xhline{2\arrayrulewidth}
		ECONIB part & Time on GPU [s]  & Time on CPU [s]  \\
		\hline
		Brain tumor segmentation  & 2     & 80    \\
		VIF region determination     & $<$1  & $<$1   \\
		Tofts model fitting  & 135   & 135    \\
		I/O, report generation        & 26    & 26     \\
		\hline
		Total                & 163   & 241    \\
		\Xhline{2\arrayrulewidth}
	\end{tabular}
\end{table}

The time required to execute all parts of ECONIB is presented in Table~\ref{tab:DCEtimeProfile}. This profile (consistent for all patients) shows that our brain tumor segmentation can be greatly accelerated ($40\times$) using a GPU. Other operations are multi-threaded and run on CPU, hence their time is the same for both hardware configurations. Overall, ECONIB requires less than 3 min and less than 4 min with and without GPU, which significantly reduces the DCE-MRI analysis time, and allows clinicians to extract reproducible results in real time.

\section{Conclusions}\label{sec:conclusions}

We introduced a fully-automated deep learning-powered approach (ECONIB) for the DCE-MRI analysis of brain tumor patients. The pivotal steps of ECONIB have been thoroughly validated using both benchmark and clinical LGG data. The experiments, backed up with statistical tests, showed that ECONIB obtains state-of-the-art reproducible results (in terms of segmentation accuracy and pharmacokinetic modeling) in a very short time which is orders of magnitude smaller when compared with other techniques. In particular, we showed that:
\begin{itemize}
  \item[-] Our deep network for brain tumor segmentation delivers very consistent and accurate segmentation for BraTS'17 and clinical data, and it allows for instant processing of full T2-weighted scans (only 2 s using a single GPU) thanks to its simplicity (it is orders of magnitude faster to train and deploy compared with top-performing BraTS segmentation engines known from the literature).
  \item[-] Our vascular input region determination delivers robust segmentation (the average root mean square error between the contrast concentration extracted from a segmented and ground-truth region is less than $3\cdot10^{-5}$) in real time (processing of full T1 VIBE scans takes less than 8 ms on average).
  \item[-] Our cubic model of the VIF yields very accurate contrast-concentration fitting. The mean square fitting error, obtained for the Quantitative Imaging Biomarkers Alliance (QIBA) phantom dataset, is an order of magnitude lower for both plasma and tissue when compared with commonly used models. Our implementation works $100\times$ faster compared to a validated state-of-the-art DCE tool.
  \item[-] Our DCE analysis pipeline requires less than 3 min for the end-to-end processing (including data loading and generating DICOM reports) using a single GPU. On a workstation which is not equipped with a GPU, the entire processing takes only 4 min (our deep learning-powered brain tumor segmentation is accelerated $40\times$ using the GPU processing when compared with its CPU version).
  \item[-] ECONIB is very flexible and can extract new quantitative DCE-MRI features (e.g.,~histogram- or texture-based).
\end{itemize}

Our current work is focused on validation of DCE biomarkers extracted by ECONIB, and comparing them with the measures obtained using other well-established software (Tissue4D, Siemens) which approximates volumes of interest by simpler geometrical objects, e.g.,~spheres (ECONIB extracts biomarkers from the very segmented volumes without any additional approximation). Once this process is finished~\cite{BARATA2017270}, ECONIB biomarkers (including texture- and histogram-based features) will help enhance diagnostic efficiency of the DCE-MRI imaging and bring new value into clinical practice.

\section*{Acknowledgements}

This work has been supported by the Polish National Centre for Research and Development under the Innomed grant POIR.01.02.00-00-0030/15.

%
%
%
%
%

\biboptions{numbers,sort&compress}
\bibliographystyle{elsart-num-sort}
\bibliography{ref_all}

%
%
%
\end{document}